\documentclass[9pt,twocolumn,twoside]{pnas-new}

\templatetype{pnasresearcharticle} 

\usepackage[utf8]{inputenc}
\usepackage{xcolor}
\usepackage{graphicx}
\usepackage{gensymb}
\usepackage{amsmath}
\usepackage{upgreek}

\title{Thermoelectric response from grain boundaries and lattice distortions in crystalline gold devices}

\author[a,1]{Charlotte I. Evans}
\author[b,1,2]{Rui Yang} 
\author[b,1]{Lucia T. Gan}
\author[c]{Mahdiyeh Abbasi}
\author[d]{Xifan Wang}
\author[b]{Rachel Traylor}
\author[b,3]{Jonathan A. Fan}
\author[a,c,d,3]{Douglas Natelson}

\affil[a]{Department of Physics and Astronomy, Rice University, Houston, TX 77005 USA}
\affil[b]{Department of Electrical Engineering, Stanford University, Stanford, CA 94305 USA}
\affil[c]{Department of Electrical and Computer Engineering, Rice University, Houston, TX 77005 USA}
\affil[d]{Department of Materials Science and NanoEngineering, Rice University, Houston, TX 77005 USA}

\leadauthor{Evans} 

\significancestatement{Scanning photothermoelectric measurements, using a laser spot as a moveable heat source, have revealed local information about thermoelectric response in various materials.  This work applies this to examine surprising thermoelectric variation in nominally simple single crystals of gold as well as individual grain boundaries between crystals.  Abrupt grain boundaries have little effect on thermoelectric response.  Instead, the Seebeck response correlates with crystallographic defects and strain associated with misorientation within the single crystals, detected via electron backscatter diffraction. Annealing reduces these thermoelectric signatures, presumably via relaxation of lattice distortions.  These measurements show that minor structural defects in otherwise single-crystalline materials can have readily detectable thermoelectric consequences, a result with implications for many devices and material systems.}

\authorcontributions{J.A.F. and D.N. conceived the experiments; C.I.E. conducted the photovoltage experiments and analyzed data; R.Y. fabricated and characterized the samples. L.T.G. conducted the EBSD measurements and analyzed the IGM data; M.A. performed the finite element simulations; X.W. assisted with photovoltage measurements; R.T. assisted with characterization measurements; C.I.E, L.T.G., J.A.F., and D.N. wrote the manuscript; All authors reviewed the manuscript.}
\authordeclaration{The authors declare no conflicts of interest.}
\equalauthors{\textsuperscript{1}C.I.E., R.Y., and L.T.G. contributed equally to this work.\\
\textsuperscript{2}Current address: University of Michigan-Shanghai Jiao Tong University Joint Institute, Shanghai Jiao Tong University, Shanghai 200240, China}
\correspondingauthor{\textsuperscript{3}To whom correspondence should be addressed. E-mail: jonfan@stanford.edu; natelson@rice.edu}


\begin{abstract}
The electronic Seebeck response in a conductor involves the energy-dependent mean free path of the charge carriers and is affected by crystal structure, scattering from boundaries and defects, and strain. Previous photothermoelectric (PTE) studies have suggested that the thermoelectric properties of polycrystalline metal nanowires are related to grain structure, though direct evidence linking crystal microstructure to the PTE response is difficult to elucidate. Here, we show that room temperature scanning PTE measurements are sensitive probes that can detect subtle changes in the local Seebeck coefficient of gold tied to the underlying defects and strain that mediate crystal deformation. This connection is revealed through a combination of scanning PTE and electron microscopy measurements of single crystal and bicrystal gold microscale devices. Unexpectedly, the photovoltage maps strongly correlate with gradually varying crystallographic misorientations detected by electron backscatter diffraction. The effects of individual grain boundaries and differing grain orientations on the PTE signal are minimal. This scanning PTE technique shows promise for identifying minor structural distortions in nanoscale materials and devices.
\end{abstract}

\dates{This manuscript was compiled on \today}
\doi{\url{www.pnas.org/cgi/doi/10.1073/pnas.XXXXXXXXXX}}

\begin{document}

\maketitle
\thispagestyle{firststyle}
\ifthenelse{\boolean{shortarticle}}{\ifthenelse{\boolean{singlecolumn}}{\abscontentformatted}{\abscontent}}{}

\dropcap{T}he thermoelectric effect, in which a conducting material develops a voltage gradient when subjected to a temperature gradient, offers a promising method for energy conversion \cite{wood_materials_1988, j.minnich_bulk_2009} and photodetection \cite{buscema_large_2013, gabor_hot_2011, lu_progress_nodate}. The Seebeck coefficient, $S$, relates the voltage gradient, $\nabla V$, generated as a response to an applied temperature gradient, $\nabla T$. When the interface of two materials with differing $S$ is heated, as in a thermocouple, an open circuit voltage develops. In metals, the electronic component of $S$ is described by the Mott-Jones equation, 
\begin{equation}
    S = - \frac{\pi^2 k_B T}{3e} \frac{d \text{ln} \sigma}{dE}\Bigr|_{E=E_F}
\end{equation}
where $\sigma$ is the energy-dependent electrical conductivity and depends on the band structure and the electron mean free path. In scanning PTE measurements, focused illumination of the sample creates a local temperature increase with a magnitude and spatial extent set by the optical absorption and thermal conduction. In wires of constant thickness and width with well-defined resistivity, the electronic mean free path should be uniform, implying that illumination far from the wire ends should therefore  generate no measurable PTE voltage. the local geometry of the wire can lead to modification of local electron and phonon surface scattering \cite{joshi_enhanced_2008, szczech_enhancement_2011, hochbaum_enhanced_2008}, producing a tailored thermoelectric response \cite{dresselhaus_new_2007}.  This manipulation of $S$ via geometric constraints has been used to produce single-metal thermocouples \cite{szakmany_single-metal_2014, sun_unexpected_2011} and is often attributed to the change of the electronic mean free path due to increased surface scattering. {The Mott-Jones relation has been used to relate this change of electronic mean free path to the absolute Seebeck coefficient as a function of film thickness in platinum thin films \cite{kockert_absolute_2019} and in Au, AuPd, and Cr/Pt films on silicon-nitride membranes \cite{mason_determining_2020}.} However, recent scanning PTE studies indicate that polycrystalline metal nanowires possessing uniform thickness and width exhibit substantial variation of the local $S$ \cite{zolotavin_substantial_2017}. Despite these thin-film devices having a uniform sheet resistance of $\sim$~1~ohms per square, far from the granular limit, $S$ varies on length scales of $\sim$~1~$\upmu$m. The variation of PTE voltage is unique to each device and changes with annealing, and changes in the magnitude of $S$ due to grain structure was suggested as a possible origin of this variation. Identifying the detailed mechanism of this variation, however, is difficult due to the structural complexity of vapor-deposited polycrystalline metal thin films.

\begin{figure*}[h!]
    \centering
    \includegraphics[width=\textwidth]{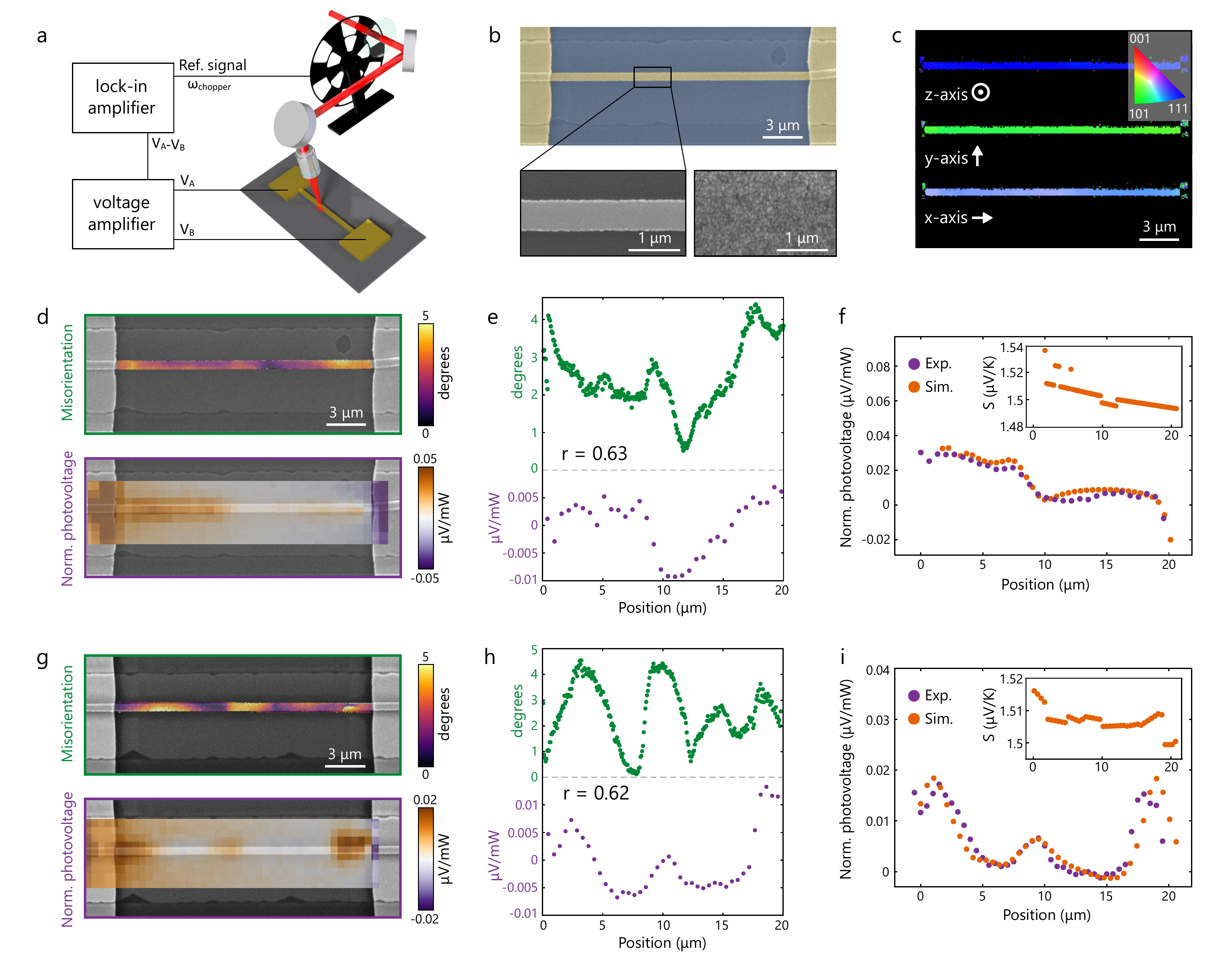}
    \caption{Single crystal gold wire. \textit{a}: Open-circuit PTE voltage is acquired as a function of laser spot position via lock-in measurements synced to laser modulation. \textit{b}: False-colored scanning electron microscope (SEM) image of a single crystal gold device on a thermal oxide substrate. Bottom left inset: zoomed-in view of a single crystal wire. Bottom right: granular structure of a polycrystalline film. \textit{c}: EBSD maps corresponding to the device in \textit{b} show the (111)-textured single crystalline nature of the device. {The EBSD maps show the local crystal directions (color coded in the inset) in the sample with respect to the z-, y-, and x-axes, respectively, which are indicated by the arrows.} \textit{d,g}: 2D maps of \textit{top}: IGM angle and \textit{bottom}: steady-state, integrated PTE voltage normalized per mW of incident laser power overlaid on SEM from \textit{b} and another representative single crystal device, respectively. \textit{e,h}: Scatter plots of \textit{top}: IGM angle and \textit{bottom}: normalized PTE voltage as a function of laser position along the length of the wire, {with a linear-in-position background subtracted to highlight the spatial variations (see SI Appendix)}. The \textit{r} value is 0.63 and 0.62, respectively, indicating a strong degree of linear correlation between the PTE voltages and IGM angles. \textit{f,i}: Scatter plot comparing the experimentally measured and simulated PTE response. Inset: Example of the local variation in Seebeck coefficient resulting in the PTE response simulated via finite element modeling.}
    \label{fig:one}
\end{figure*}

Recent breakthroughs in the top-down fabrication of crystalline gold microstructures provide a new means for studying the thermoelectric properties of metallic systems at the one to few grain level. These structures are prepared via the metal-on-insulator rapid melt growth process, which enables the direct preparation of high-quality single crystal and bicrystal gold wires, consisting of two grains separated by an individual grain boundary, on thermally and electrically insulating amorphous oxide substrates \cite{zhang_single-crystal_2018,gan_high-throughput_2019}. These structures are lithographically patterned into wires and can be defined with single grain boundaries that bisect each wire. Leveraging this novel growth technique allows us to probe local thermoelectric behavior in model material systems that are grain boundary-free or possess a single grain boundary. 

Here, we establish an unexpected strong connection between the photothermoelectric response and crystalline defects associated with small degrees of lattice curvature in single crystal and bicrystal gold wires as probed by electron backscatter diffraction (EBSD). The PTE photovoltage signal of the single crystalline wires varies on the same length scales as the long-range intragranular misorientations. The PTE response at individual grain boundaries is undetectably small, demonstrating that $S$ is much more sensitive to local geometric changes and microscale orientation variations than to abrupt single plane defects like grain boundaries. Annealed devices demonstrate a 2.7$\times$ decrease in PTE response compared to unannealed devices, further confirming that distributions of dislocations and strain associated with microscale orientation variations beyond changes to granular structure can play a considerable role in the thermoelectric response of micro- and nanoscale devices. While uniform tuning of Seebeck response had previously been examined in macroscopic, polycrystalline wires \cite{mortlock_effect_1953, amuzu_effect_1981, kleber_sensitivity_2009}, the present measurements provide a new perspective on local variations of $S$ due to defects, dislocations, and strain. 

\section*{Results and Discussion}

We employ the scanning PTE microscope setup in Fig. \ref{fig:one}a to probe the local thermoelectric response. The laser spot acts as a scannable heat source. Since the spot size is larger than the wire width, we can approximate the wire as 1D, with the spot generating a temperature profile $T(x)$, where the $x$-direction is along the long axis of the wire. The PTE photovoltage reflects the integrated response of the spatially dependent $S(x)$ and $T(x)$. The steady-state, integrated PTE open circuit voltage of each device is measured as a function of laser position via lock-in techniques at room temperature under high vacuum, and it uses a 785~nm diode laser modulated by an optical chopper. The position of the beam is raster-scanned using a piezo-controlled lens before being focused onto a device by an objective to a spot with a full-width half-maximum (FWHM) of 1.8~$\upmu$m. The PTE voltage measurements throughout this work are normalized to the incident laser power of 15~mW. The linear power dependence of the PTE signal at a stationary location is shown in SI Appendix, Fig. S1. The samples are 100~nm thick and 600~nm wide gold single crystal or bicrystal microscale wires grown on 300~nm thermal oxide substrates with deposited contact electrodes. More information regarding the device fabrication and PTE measurements can be found in the Methods. The single crystal wires are 20~$\upmu$m in length, shown in the SEM image in Fig. \ref{fig:one}b. As seen in the bottom left inset of the SEM in Fig. \ref{fig:one}b, the single crystal wire is smooth compared to the granular structure of a similarly thick polycrystalline film on the bottom right. The corresponding EBSD maps display a (111)-textured single crystal wire (Fig. \ref{fig:one}c), consistent with previous work \cite{zhang_single-crystal_2018}. {EBSD is a SEM-based technique in which backscattered electrons form a diffraction pattern that provides information about the local crystallographic structure of the imaged sample. The local crystal orientation (as shown in the color scale of the inset to Fig.~\ref{fig:one}c) is determined with respect to the x-, y-, and z-axis of the sample as indicated in Fig. \ref{fig:one}c by the arrows.}

The bottom panel of Fig. \ref{fig:one}d shows the 2D PTE photovoltage map corresponding to the single crystalline wire in Fig. \ref{fig:one}b-c. The largest signal occurs at the interface of the contact pads and the wire, with each pad having similar voltage magnitudes but opposite polarity, consistent with the definition of the open circuit voltage configuration shown in Fig. \ref{fig:one}a. This is expected, as the interface of the pads and the wire should act as thermocouples due to the differences in $S$ caused by a change in metal thickness and increased scattering at the contacts \cite{sun_unexpected_2011, szakmany_single-metal_2014, dresselhaus_new_2007}. These contacts between the wire and the pads at each wire end introduce a small linear-in-position background in the PTE signal. In addition, the morphologically featureless device displayed in the inset to Fig. \ref{fig:one}b produces small PTE signals that vary along the length of the wire on length scales comparable to the laser spot size (for reference, the pixel length is $\sim$~0.5~$\upmu$m, approximately 4$\times$ smaller than the laser FWHM). The variations in the PTE voltage indicate that the value of $S$ changes along the length of the wire despite the uniform resistivity and the absence of any geometric variation, a regime where $S$ would ordinarily be expected to be constant. 

Structural character more subtle than the gross morphology (e.g. obvious defects, mechanical deformations) must be responsible for local variations in $S(x)$.  Rotations in the crystalline structure can be observed by further analyzing the local misorientation of the crystal from the EBSD data in Fig. \ref{fig:one}c. Analysis of the microstructure of the underlying gold samples is enabled through large-area EBSD scans \cite{dingley1997electron}. The orientation data taken from the diffraction patterns are used to construct 2D intragranular misorientation (IGM) maps to uncover small changes between neighboring points in the crystal \cite{wright_review_2011,brewer_misorientation_2002}. IGM is a pixel-based measure of the orientation spread in a grain, which is computed from the misorientation between each pixel and the grain-averaged orientation. This technique maps the microscale orientation gradients that manifest in single crystals, displayed in the top panel of Fig. \ref{fig:one}d. The samples appear to be single crystalline with long range imperfections over tens of microns with subtle low angle lattice curvatures about the $\langle111\rangle$ axis of up to 5 degrees of angular deviation from the average orientation. Comparing the IGM and PTE maps of the same devices enables the systematic examination of locations with photovoltage signal variation and crystal deformation in the same devices. 

The link between the IGM and PTE signal is clear in the scatter plots in Fig. \ref{fig:one}e after averaging the signal in the 2D maps in Fig. \ref{fig:one}d along the width of the wire. The linear-in-position PTE background mentioned above is subtracted from the PTE measurements to highlight deviations of the PTE signal. This background subtraction does not qualitatively change the results and enables correlation statistics (raw data displayed in SI Appendix, Fig. S2). The scatter plots of the two measurements show variations in signal in the same location with similar length scales, despite the resolution of the IGM measurements being much finer than the laser spot size. The Pearson correlation coefficient, \textit{r}, was computed to determine the linear correlation between the PTE and IGM values using interpolated PTE values at even intervals determined by spline fitting. The \textit{r} value of the data set in Fig. \ref{fig:one}e is 0.63, indicating a strong degree of linear correlation. This agreement between the PTE voltages and IGM values is seen consistently. Fig. \ref{fig:one}g-h shows the results of another single crystal device with an \textit{r} of 0.62. 

The IGM measurements probe orientation variations that arise from sources including disclinations, dislocations, stacking faults, and strain along the length of the wire, demonstrating that the photovoltages are sensitive to defect accumulations that mediate low angle lattice rotations. Due to the (111)-texturing normal to the substrate, gradually varying orientation gradients manifest as very localized low angle tilt boundaries in plane formed by arrays of dislocations \cite{cai2016imperfections}. These low angle rotations appear through the emergence and multiplication of dislocations that are accompanied by their own elastic fields. While the atomic structure cannot be resolved with a scanning electron beam, we measure the resultant angular deviations as a proxy for the presence of a combination of strain and defects. Strain and dislocation density in this system cannot be decoupled, but the combined effects affect the carrier transport properties that result in changes in $S$ that can be detected via the PTE voltage technique.

With finite element modeling, it is possible to model the changes in Seebeck coefficient required to produce the observed changes in photovoltage magnitude as a function of laser position, shown in Fig. \ref{fig:one}f and \ref{fig:one}i. We perform modeling using COMSOL Multiphysics Joule Heating Physics, with details described in the SI Appendix. We assign a spatially varying $S(x)$ along the length of the wire, and good thermal contact with the substrate and standard optical properties are assumed within the gold wire. A Gaussian heat source with the same FWHM of the laser beam is applied to the surface and the corresponding voltage is probed. The insets of Fig. \ref{fig:one}f and \ref{fig:one}i show one particular distribution of $S$ that can provide the variation of photovoltages observed in experiment, varying $\sim~0.2\%$ over 1~$\upmu$m. Although there is not a single unique solution for the spatial variation in $S$, the simulations provide insight on the overall magnitudes of changes in $S$ required to produce the magnitude and length scales of the observed PTE voltages.  For a sense of scale, measurements on bulk polycrystalline gold wires imply that tensile strain of 100\% would change the bulk $S$ by 6.3~$\upmu$V/K \cite{amuzu_effect_1981}. The  small changes in $S$ inferred from the simulations would then correspond to effective local tensile strains of $\sim$~0.15\%, though caution is warranted in any comparison with macroscopic measurements on polycrystalline wires.

\begin{figure*}[h!]
    \centering
    \includegraphics[width=\linewidth]{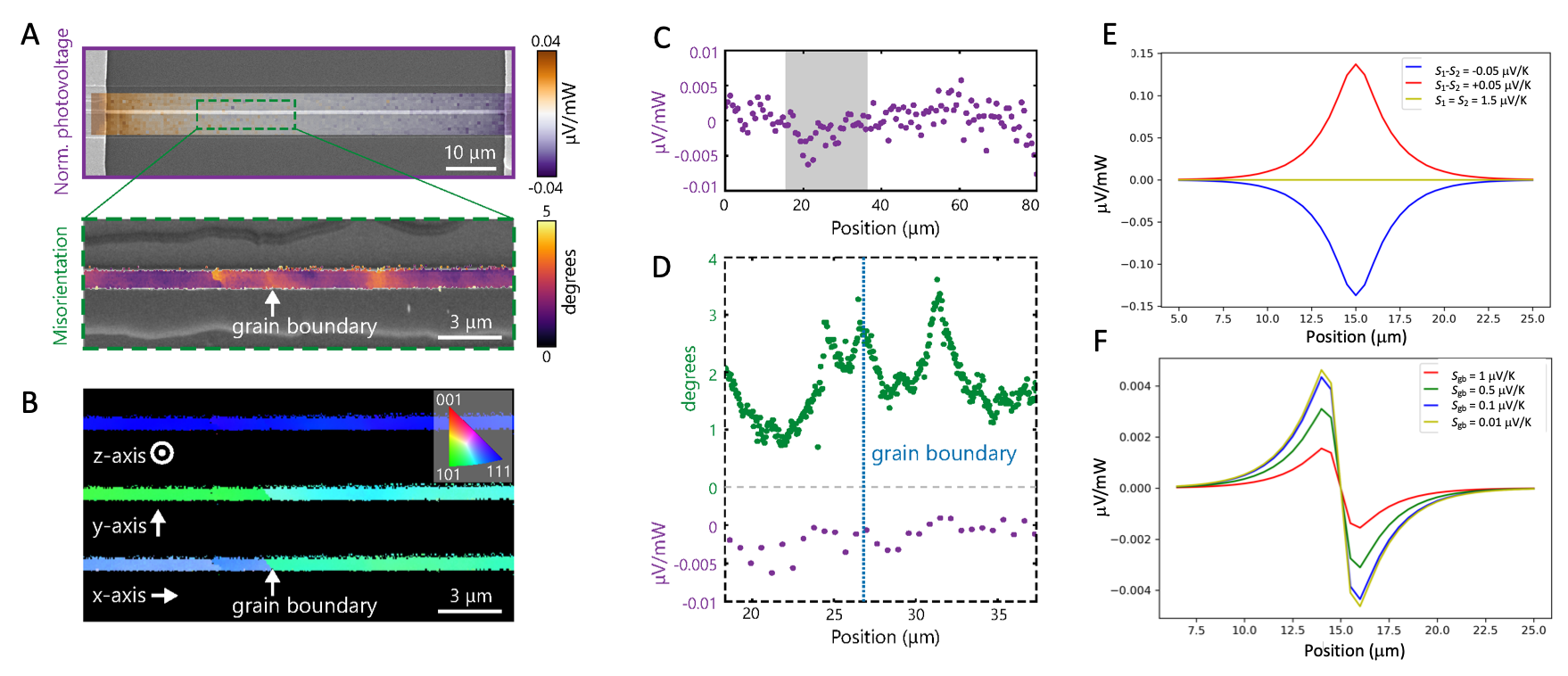}
    \caption{Annealed bicrystal gold wire. \textit{a}: 2D map of normalized PTE signal superimposed on a SEM image of a bicrystal device. Inset: 2D IGM map overlaid on zoomed-in SEM image. \textit{b}: EBSD maps corresponding to the green dashed box in \textit{a}. The bicrystal is (111)-textured in the $z$-axis and the $x$- and $y$-axes show a clear grain boundary. \textit{c}: Scatter plot of the normalized PTE signal along the length of the device. The grey shaded area corresponds to the green dashed box in \textit{a}. \textit{d}: Scatter plots of PTE and IGM signal from the green dashed box in \textit{a}. The grain boundary is marked by the blue dashed line.  {\textit{e}: Examples of simulated PTE profiles for an abrupt simple junction between metals of identical thermal conductivities but differing Seebeck coefficients $S_{1}$ and $S_{2}$ (see SI Appendix for greater detail).  \textit{f}: Examples of simulated PTE profiles for a small segment of “grain boundary” with Seebeck coefficient $S_{gb}$ sandwiched between extended regions with $S = 1.5 \mu$V/K (see SI Appendix for greater detail)}.
    }
    \label{fig:two}
\end{figure*}

To study the effect of individual grain boundaries on PTE signal, we examine bicrystal devices consisting of two grains bisected by a single grain boundary, approximately 80~$\upmu$m in length. In order to highlight signal from the grain structure and reduce residual strain from fabrication, the devices were annealed at 400\degree C for 1 hour prior to measurement. Due to the length of the bicrystal devices, detailed IGM measurements are taken in a 20~$\upmu$m window around the grain boundary, and each grain is analyzed independently against its own average orientation reference. Fig. \ref{fig:two}a shows an SEM image of a bicrystal device overlaid with a 2D PTE voltage map, which has relatively little signal along the length of the device. The green box indicates the window where we perform the EBSD and IGM analyses. The IGM map overlaid on the SEM show there are still subtle variations in orientation around the grain boundary despite the lack of notable PTE voltage around the same area (Fig. \ref{fig:two}a inset). Fig. \ref{fig:two}b shows the EBSD maps of this device, with a $\langle 111 \rangle$ orientation in the $z$-axis normal to the substrate and a distinct grain boundary evident in the $x$- and $y$-axis maps, consistent with previous work \cite{gan_high-throughput_2019}. Fig. \ref{fig:two}c shows the almost featureless photovoltage signal along the length of the wire, with scatter limited by the measurement noise floor. 

Fig. \ref{fig:two}d compares the IGM and photovoltage measurements around the grain boundary, which is marked by the blue dashed line. The IGM data shows a small degree of change in misorientation around the grain boundary, whereas the PTE photovoltage remains relatively low in signal, showing that individual abrupt grain boundaries do not behave detectably as thermocouples relative to the long-range misorientation. Finite element simulations detail the change of $S$ required to observe the PTE voltages observed in experiment (SI Appendix, Fig. S15), which implies that different grain orientations must have $S$ values that differ by less than 0.013\% (0.0002~$\upmu$V/K) over 500~nm. Although the PTE response due to grain boundaries is negligible, the measurement is sensitive to long-range misorientation changes on either side of the boundary (SI Appendix, Fig. S3). These observations imply that the spatially varying $S(x)$ responsible for the complex PTE maps seen in polycrystalline wires \cite{zolotavin_substantial_2017} arises from a complex combination of geometry, strain, and dislocation formation, rather than the grain boundaries themselves. {Note that an abrupt change in Seebeck coefficient would manifest itself as a peak in the PTE voltage as a function of laser position (Fig. 2e), as expected for a simple thermocouple heated at the junction between dissimilar materials.  An inclusion of a small (relative to spot size) region of different S than the surrounding material creates back-to-back thermocouples, leading to an antisymmetric PTE feature as a function of laser position (Fig. 2f).  See SI Appendix for greater detail of these simulations.} 

\begin{figure*}[h!]
    \centering
    \includegraphics[width=\linewidth]{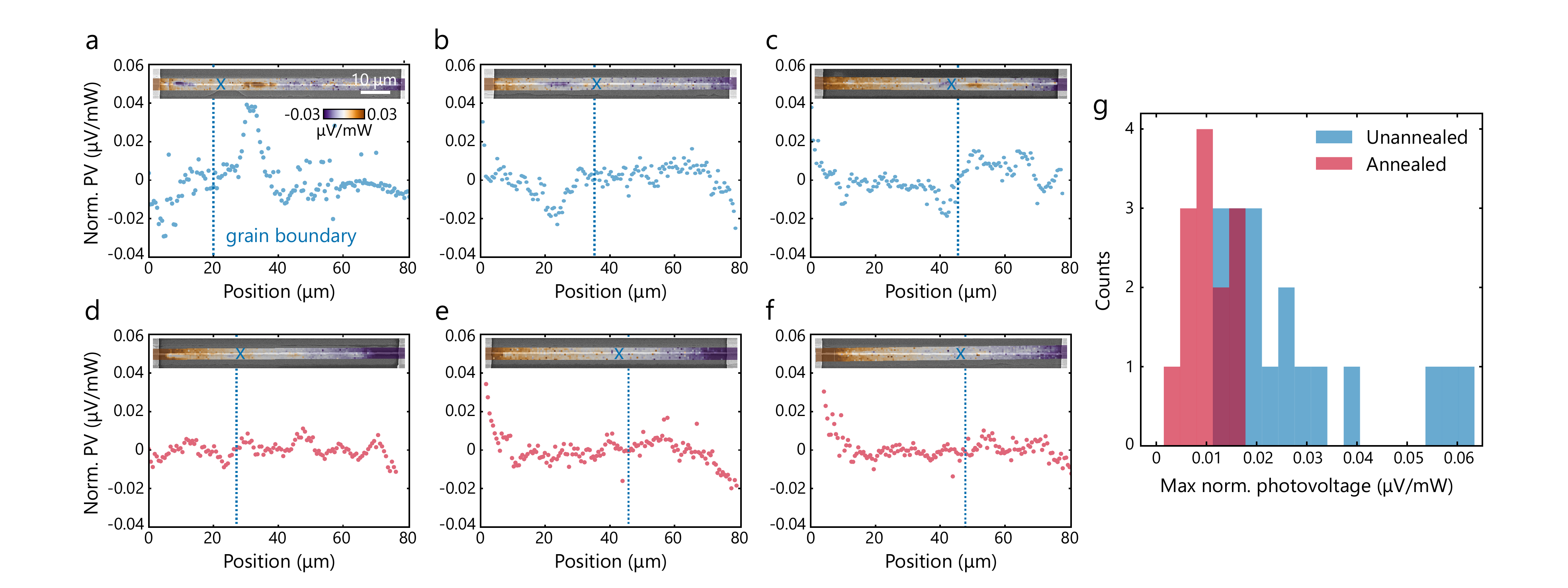}
    \caption{Comparison of unannealed and annealed bicrystals: \textit{a-c}: Normalized PTE voltage maps and scatter plots of three unannealed bicrystals. \textit{d-f}: PTE voltage maps and scatter plots of three annealed bicrystals. On average, the PTE voltage profiles of the unannealed devices had greater spatial variation and larger magnitudes. \textit{g}: Histogram of maximum magnitude of signal in 19 unannealed (blue) and 13 annealed (red) bicrystal devices.}
    \label{fig:three}
\end{figure*}

Unannealed bicrystals typically have larger PTE responses than annealed devices. The top and bottom rows of Fig. \ref{fig:three} compare the PTE voltage profiles of unannealed and annealed devices, respectively. The annealed devices were annealed at 400\degree C for 1 hour prior to measurement, as mentioned previously. Due to fragility of the wires arising from the lack of an adhesion layer between the crystals and substrate, the devices were either annealed or left unannealed prior to wire bonding on the chip carrier; the comparison is of unique devices. The profiles of the unannealed devices have, on average, greater spatial variation and signal magnitude than the annealed devices. In all of these cases, no notable change in PTE signal occurs at the grain boundary, marked by blue crosses in the maps and by blue dashed lines in the scatter plots. The maximum signal along the length of 19 unannealed and 13 annealed bicrystal devices of comparable dimensions and no gross morphological defects are compared in a histogram in Fig. \ref{fig:three}g. The unannealed devices clearly contain larger PTE magnitudes than the annealed devices. Unannealed bicrystal devices have an average signal of $0.026 \pm$ 0.006 $\upmu$V/mW and annealed devices have an average signal of $0.0096 \pm$ 0.0003 $\upmu$V/mW. On average, unannealed bicrystals have a $2.7\times$ larger signal than the annealed ones, demonstrating that defect annihilation and strain relaxation during annealing can contribute considerably to the local $S$. We note that annealing at 400\degree C does not modify the crystalline character sufficiently to produce large detectable changes in the IGM maps. Previous analyses suggest that during annealing, dislocations rearrange and coalesce to form stable arrays that reduce the dislocation density over long ranges \cite{forty1954direct}, which could be the source of lower PTE magnitudes. This implies that the PTE response arises from a combination of crystallographic defects, only some of which manifest themselves in the IGM signal.

Additionally, consistent with prior work on single-metal thermocouples, the PTE technique can readily detect changes in $S$ due to local geometric changes associated with changing metal thickness or width.  Prominent morphological defects can overwhelm the more subtle structural effects detected above. SI Appendix Fig. S6 depicts two crystalline wires with different morphological defects. Illuminating the defects result in signature changes in the PTE signal, which describes the change in Seebeck coefficient (SI Appendix Fig. S10 and S11). SI Appendix Fig. S7a-b shows the SEM and EBSD images, respectively, of a polycrystalline wire with a large defect in device thickness. Although there are considerable changes in the granular structure of the device, the PTE response is dominated by the defect and no variation is observed due to the grain boundaries or intragranular features (SI Appendix Fig. S7c-d). Simulations, discussed at length in the SI Appendix, suggest that the change in $S(x)$ due to this type of geometric defect is significantly larger than changes in $S(x)$ due to subtle defects of crystalline structure. Changes of the geometric structure can be readily detected via photovoltage measurements and provide signatures that depend on the details of the structural change. While structural defects \textit{can} dominate the photovoltage signal, our measurements show that crystallographic misorientation and deformation can result in photovoltages of detectable magnitudes. This insight is only possible by having a clean model system such as that provided by the single crystal and bicrystal structures grown by rapid melt growth.

Scanning PTE microscopy is a powerful tool that can provide insight on structural and electronic features within a microscale device. Performing measurements on single crystal and bicrystal wires reveal nontrivial dependences of $S$ on grain microstructure that are critical to understand in nanostructured thermoelectric materials.  Mapping the PTE photovoltages can indicate locations where there are gradually varying misorientations, which necessarily indicates locations of strain and dislocations. Strong correlations exist between the measured photovoltages across single crystal gold structures and corresponding intragranular misorientation profiles measured by EBSD.  The analogous measurements performed on bicrystal gold structures reveal that abrupt higher angle grain boundaries do not noticeably affect $S(x)$. Annealed bicrystal devices show that the overall photovoltage response associated with misorientation is reduced by a factor of 2.7, again highlighting that changes in dislocation density and strain affect local thermoelectric response. The present results indicate that thermoelectric measurements of nanostructured materials should carefully consider the underlying crystalline microstructure and strain as potential contributors of signal and device-to-device variation. 

\matmethods{

\subsection*{Fabrication}
Gold single crystal and bicrystal wires were synthesized on a thermally oxidized silicon substrate using the rapid melt growth process, detailed in previous publications \cite{zhang_single-crystal_2018, gan_high-throughput_2019}. 100~nm thick vapor-deposited gold microstructures along with 30~nm thick platinum seeds were defined through a photolithograpy and lift-off process on 300~nm of thermal oxide. The metal system was encapsulated in an insulating low-pressure chemical vapor deposition (LPCVD) silicon dioxide crucible and heated in a rapid thermal annealing system to 1080\degree C, above gold's melting point (1064\degree C), at a rate of 15\degree C/s. After holding at 1080\degree C for 1 second, the metal system was cooled at a similar rate. During cooling, liquid epitaxial growth drove single crystal solidification that nucleated from each seed. Microstructures seeded at one end yielded single crystals, while those seeded from both ends yielded bicrystals that were each bisected by a $\langle 111 \rangle$ tilt grain boundary. The insulating crucible was removed through dry plasma etching. Wires with a width of about 600~nm were defined by means of electron beam lithography and ion milling on the gold crystals. Gold electrical contact pads 150~nm thick with a Ti adhesion layer were deposited on each wire to define single crystal and bicrystal wires that were 20~$\upmu$m and 80~$\upmu$m in length, respectively. Each grain was (111)-textured normal to the substrate with different in-plane crystal orientations ranging between $\langle 110 \rangle$ and $\langle 112 \rangle$, and each bicrystal possessed a $\langle 111 \rangle$ tilt grain boundary, consistent with previous works. The chip was then mounted on a chip carrier and wire bonded for photovoltage measurements.

\subsection*{Experiment}
A home-built scanning laser microscope was used to optically heat the device kept in a closed cycle cryostat (Montana Instruments) under high vacuum at room temperature. A linearly polarized 785~nm CW laser diode with a maximum incident intensity of 700~mW/cm$^2$ was focused to a spot with FWHM diameter of 1.8~$\upmu$m. The intensity of the laser was varied using neutral density filters. The laser polarization was rotated using a half wave plate to 90\degree, perpendicular to the length of the wire, unless specifically mentioned. A mechanical chopper was used to modulate the incident laser intensity to a square wave of frequency 287~Hz. The chopper was sufficiently slow compared to the thermalization timescale that the measured voltage is a steady-state property \cite{benner_lateral_2014}. The chopper frequency was used as the external reference for the SR7270 DSP lock-in amplifier. The open circuit voltage of the device was detected by measuring the amplified potential difference between the two ends using a SR560 low noise precurrent amplifier as the input to the lock-in amplifier. A total of 55 single crystals and 40 bicrystals were measured. The devices discussed in this work {are examples} of typical device behavior. {The statistical uncertainties in the individual PTE data points as plotted in the maps are small (comparable to the sizes of the data markers in the cross-section plots).  The dominant uncertainties are systematic, associated with small hysteresis in the scanning stage (such that the locations of particular pixels relative to the sample can vary slightly from scan to scan), the stitching of separately scanned fields of view together, and possibly small variations in depth of focus of the laser.}

SEM and EBSD maps were collected using a FEI Magellan 400 XHR and a Thermo Fisher Scientific Apreo S LoVac scanning electron microscope equipped with a Bruker Quantax e-Flash(HR) EBSD detector. {EBSD maps were collected using an accelerating voltage of 25 kV at an incident angle of 70\degree from the substrate normal.} The orientation data taken from the diffraction patterns was used to construct 2D intragranular misorientation (IGM) maps. EBSD orientation data was imported and analyzed using MTEX in MATLAB. The average orientation of each grain was computed and used as a reference orientation. The misorientation between a pixel's orientation and the reference was computed for each pixel to yield the IGM maps. Due to the face-centred cubic (fcc) symmetry of gold, the misorientation is always a positive value. The pixel size of the EBSD and IGM maps was approximately 90~nm. 

\subsection*{Simulations and modeling}
Modeling was performed using COMSOL Multiphysics versions 5.3a and 5.4. The PTE measurements were implemented using Joule Heating physics. Known optical properties for gold were used to calculate expected absorption of the laser spot.  An equivalent Gaussian heat source was then applied to the modeled device structure, with appropriate thermal material properties and boundary conditions to find $T(x)$ for a given laser spot position. Further details are described in the SI Appendix.

\subsection*{Data availability}
{All data supporting the findings of this study are available in the manuscript, the SI, and/or from a public archive (https://dx.doi.org/10.5281/zenodo.3901826).}

}

\showmatmethods{} 

\acknow{The authors thank Kevin Kelly for helpful conversations and William D. Nix for contributing to the interpretation of the results.

D.N. and C.I.E. acknowledge support from Robert A. Welch Foundation grant C-1636.  D.N., M.A., and X.W. acknowledge support from National Science Foundation Award No. ECCS-1704625. J.A.F., L.T.G., R.Y., and R.T. acknowledge support from the National Science Foundation under Award 1804224, the Air Force Office of Scientific Research Multidisciplinary University Research Initiative (MURI) under Award No. FA9550-16-1-0031, and the Packard Fellowship Foundation. R.Y. acknowledges support from Shanghai Sailing Program under Award 19YF1424900, and from University of Michigan–Shanghai Jiao Tong University Joint Institute in Shanghai Jiao Tong University. Part of this work was performed at the Stanford Nanofabrication Facility and the Stanford Nano Shared Facilities, supported by the National Science Foundation under Award ECCS-1542152.}

\showacknow{} 

\bibliography{sample}

\end{document}